\documentclass[useAMS,usenatbib]{mn2e}
\usepackage{graphicx}
\title[Intrinsic Shapes of Very Flat Elliptical Galaxies]{Intrinsic Shapes of Very Flat Elliptical Galaxies}
\author[D. K. Chakraborty, A. K. Diwakar and S. K. Pandey]{D. K. Chakraborty$^{1}$\thanks{E-mail: chakrabortydeokumar@gmail.com}, A. K. Diwakar$^{1}$\thanks{E-mail: diw.arun@gmail.com} and S. K. Pandey$^{1}$\thanks{E-mail: skp@iucaa.ernet.in}\\
$^{1}$ School of Studies in Physics \& Astrophysics, Pt. Ravishankar Shukla University, Raipur, Chhattisgarh 492010, India}

\begin{document}

\date{Accepted xx. Received xx; in original form xx}

\pagerange{\pageref{firstpage}--\pageref{lastpage}} \pubyear{2010}

\maketitle

\label{firstpage}

\begin{abstract}
Photometric data from the literature is combined with triaxial mass models to derive variation in the intrinsic shapes of the light distribution of elliptical galaxies $NGC 720, 2768$ and $3605$. The inferred shape variation in given by a Bayesian probability distribution, assuming a uniform prior. The likelihood of obtaining the data is calculated by using ensemble of triaxial models. We apply the method to infer the shape variation of a galaxy, using the ellipticities and the difference in the position angles at two suitably chosen points from the profiles of the photometric data. Best constrained shape parameters are found to be the short to long axial ratios at small and large radii, and the absolute values of the triaxiallity difference between these radii.

The elliptical galaxies of our present investigation are very flat, with ellipticity typically around $0.3$ or more. We find that the expectation values of the short to long axial ratio of these galaxies are around $0.5$.
\end{abstract}


\section{Introduction}
Intrinsic shapes of the individual elliptical galaxies have been investigated by \citet{b2}, \citet{b22}, \citet{b17,b18}, \citet{b1}, \citet{b19}, and \citet{b21}. These authors have used the kinematical data and the photometric data, and have used the triaxial models with the density distribution $\rho(m^{2})$, where $m^{2}=x^{2}+y^{2}/p^{2}+z^{2}/q^{2}$ with axial ratios $p$ and $q$. Here, $(x, y, z)$ are the usual Cartesian co-ordinates, oriented such that $x$-axis ($z$-axis) lies along the longest (the shortest) axis of the model. It was shown analytically that the projected density of such a distribution $\rho (m^{2})$ with constant $(p, q)$ is stratified on similar and co-aligned ellipses \citep{b16,b2}. \citet{b17} uses (apart from the kinematical data) a constant value of ellipticity, which is an average over a suitably chosen range of radial distance, for the shape estimates. The shape estimates are robust, and are described by a pair of the shape parameters, namely the short to long axial ratio $c_{L}$ of the light distribution and the triaxiality $T_{M}$ of the mass distribution.

A complementary problem was attempted by (\citealt{b5}, hereafter $C08$), wherein variation in the intrinsic shapes of the light distribution of elliptical galaxies was investigated by using triaxial models, which exhibit ellipticity variation and position angle twist. These models are fixed by assigning the values of axial ratios $(p_{0}, q_{0})$ and $(p_{\infty}, q_{\infty})$ at small and at large radii, respectively. These axial ratios are related to triaxialities $T_{0}$ and $T_{\infty}$, respectively, at small and large radii. We use Bayesian statistics, and obtain the variation in the shape, following the methodology described in \citet{b17}. We find that the marginal posterior density $(MPD)$ is likelihood dominated, so that it relatively insensitive to the unknown prior density. We use a flat prior. We use a large ensemble of models, so that the shape estimates may be model independent.

The basic ingredients of our method are the same as in \citet{b17}, and we adopt all the necessary alterations described in $C08$. We use $(q_{0}, T_{0}, q_{\infty}, T_{\infty})$ as the shape parameters and use the ellipticities $\epsilon_{in}, \epsilon_{out}$ and the position angle difference $\Theta_{out}-\Theta_{in}$ at two suitably chosen points $R_{in}$ and $R_{out}$ from the profiles of the photometric data of the galaxies. We find that the best constrained shape parameters are $q_{0}, q_{\infty}$ and the absolute value of the triaxiality difference $T_{d}$, defined as $|T_{d}| = |T_{\infty}-T_{0}|$. 

$C08$ have estimated the shapes of $10$ elliptical galaxies which are comparatively rounder, with ellipticities $\le 0.3$. We now investigate shapes of three more galaxies, namely $NGC 720, 2768$ and $3605$. These are very flat galaxies with ellipticity around $0.3$ or more. We find that the expectation values of the short to long axial ratio of these galaxies are around $0.5$. We use triaxial models which are very flat. We take models with the lower limit of $(q_{0}, q_{\infty}) \sim  0.3$ for our shape investigation. We find that a class of very flat triaxial models develops several undesirable features (sect. 2 and Appendix A), and are not employed in the present shape estimates.

Determination of the intrinsic shape using photometry is important because the number of galaxies with good photometric is many more than those with good kinematics. Besides, the results obtained by alternative models and techniques can be used for a comparison. Photometry constrains the flattening $(q_{0}, q_{\infty})$ but can not constrain $(T_{0}, T_{\infty})$. Thus, our work is complementary and not contradictory to that of Statler and his coworkers.

Sect. 2 presents the models. The necessity for the choice of small values of the lower limits of $(q_{0}, q_{\infty})$, and the intrinsic shapes of the galaxies are presented in sect. 3. Sect. 4 is devoted to results and a discussion. 

\section{Model}
We use models, which are triaxial generalizations of the spherical $\gamma $ models of \citet{b7}, with density $\rho$ given by
\begin{eqnarray}
\rho(r) = \frac{M_{0}(3-\gamma)b}{4\pi}r^{-\gamma}(b+r)^{-4+\gamma},
\end{eqnarray}
where $M_{0}$ is the mass of the model, $r$ is the radial coordinate, $0\leq \gamma <3$ and $b$ is the scale length. The models have cusp at the centre, and the density decreases as $r^{-4}$ at large radii. Dehnen's models are the generalization of the well studied models of \citet{jaf83} and \citet{her90}, corresponding to $\gamma = 2$ and $\gamma = 1$, respectively. The projected surface density of the model of Dehnen, corresponding to $\gamma = 1.5$, most closely resembles to the de Vaucouleurs $R^{1/4}$ law. Presently, we concentrate to $\gamma = 1.5$ models only.

A triaxial generalization of $(1)$ is presented in \citet{b3}, which is modified in $C08$. The model is the density distribution of the same form as (1) with $r$ replaced by $M$, where
\begin{eqnarray}
M^{2} = x^{2}+\frac{y^{2}}{P^{2}}+\frac{z^{2}}{Q^{2}} \ ,
\end{eqnarray}
with varying axial ratios
\begin{eqnarray}
P^{-2}(M) = \frac{\beta b^{2}p_{0}^{-2}+M^{2}p^{-2}_{\infty}}{\beta b^{2}+M^{2}} \ , 
\end{eqnarray}
and
\begin{eqnarray}
Q^{-2}(M) = \frac{\beta b^{2}q_{0}^{-2}+M^{2}q^{-2}_{\infty}}{\beta b^{2}+M^{2}} \ .
\end{eqnarray}
The axial ratios $(P,Q)$ reduce to $(p_{0},q_{0})$ at small radii and to $(p_{\infty},q_{\infty})$ at large radii. $\beta >0$ is a parameter, which for a choice of $(p_{0},q_{0},p_{\infty},q_{\infty})$ alters $P$ and $Q$ in the intermediate region. The models are fixed, once the axial ratios $(p_{0}, q_{0}, p_{\infty}, q_{\infty})$ are chosen. The triaxialities $T_{0}$ and $T_{\infty}$ are related to the axial ratios at small and at large radii by
\begin{eqnarray}
T_{0} = \frac{1-p_{0}^{2}}{1-q_{0}^{2}}. \quad T_{\infty} = \frac{1-p_{\infty}^{2}}{1-q_{\infty}^{2}} \ .
\end{eqnarray}
To fix up the scale length $b$ of the triaxial models, we consider $\gamma = 1.5$ and use the value of the effective radius $R_{e} = 1.28b$ of the spherical model. The effective radius of the triaxial models depends on the axial ratios, as well as on the viewing angles. However, such changes are small for $\gamma$ models \citep{b10}, and are neglected.

The constant $\rho$ surfaces are coaxial ellipsoids. Projection of these models on a plane perpendicular to a line of sight, and therefore, the calculation of ellipticity and position angle are performed numerically. We refer to these models as $M^{2}$ models.

Another form of triaxial generalization of $(1)$ is investigated by \citet{b10}, where two more terms are added to equation $(1)$, each one of these is a suitable radial function multiplied by spherical harmonics of low order. The models provide simple analytical representation of the observed surface brightness of triaxial elliptical galaxies. However, for large values of flattening models become peanut shaped and are not used in the present investigation. Very flat de Zeeuw - Carollo models are discussed in Appendix A.

\section{Intrinsic shapes}
The galaxies chosen here are very flat. The morphological classification of $NGC720$, $2768$ and $3605$ are $E5/E3, E6/E5$ and $E4/E5$ respectively, from $RC2$ \citep{b9} catalogue. The apparent flattening of a elliptical galaxy depends on the intrinsic flattening and the orientation. Further, the marginal posterior density $(MPD)$ ${\cal{P}}$ of the Bayesian estimate is obtained by integrating the posterior density over all viewing angles. To gain some insight into the possible values of the intrinsic shape, which will be obtained by Bayesian method, we perform the following numerical experiments. The objective of these experiments is to find suitable limits of the axial ratios $(q_{0}, q_{\infty})$ for the plots of ${\cal{P}}$. In the plots of ${\cal{P}}$ in $C08$, $q_{0}$ and $q_{\infty}$ extend from $0.5$ to $ 1.0$. \citet{b17} chooses $0.4 \le c_{L} \le 1.0$.

Fig. 1 shows the plot  between the number $N$ of viewing angles $(\theta^{'}, \phi^{'})$ and the axial ratio $q$ which gives ellipticity $0.50 \le \epsilon \le 0.55$ (plot 1A) and $0.07 \le \epsilon \le 0.17$ (plot 1B). The number of viewing angles is counted between $0^{o}.0$ and $90^{o}.0$ at the interval of $1^{o}.0$, both for $\theta^{'}$ and $\phi^{'}$. The total number of viewing angles in this numerical experiment is $8100$. The axial ratio $p$ is taken as $0.9$. Here, we use Stark model. We find that a higher values of ellipticity is produced by flatter models and a lower values of ellipticity is produced by rounder models, over a larger number of viewing angles. Therefore, the Bayesian estimate should pick up a flat model to represent shape of the galaxies of our present investigation. It is interesting to note that the plots $(1A \& 1B)$ show a maxima, which lies at $q \sim 0.44$ for the plot $1A$ and at $q \sim 0.82$ for the plot $1B$

The fig. 1 and its inferences are based on applying Stark model, which has constant values of the axial ratio $(p,q)$. However, in our shape estimates, we use models with varying axial ratios. So, we re-examine the results of these plots by considering $M^{2}$ models.

\begin{figure}
\includegraphics[width=8.0cm]{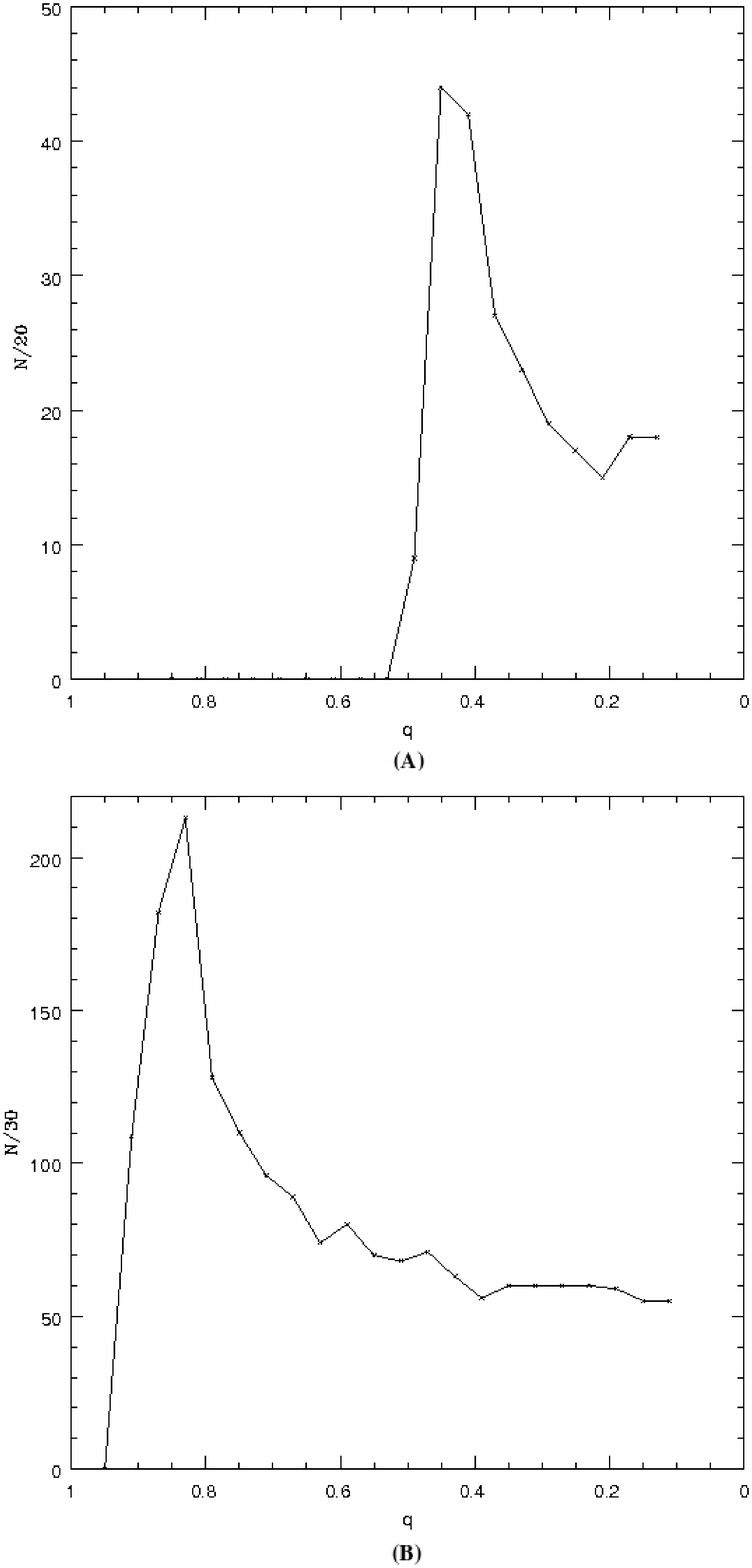}
\caption{\small Plot between the number of viewing angles and the axial ratio $q$, which would reproduce  ellipticity in a chosen interval. Figure 1A is drawn for the ellipticity $0.50 \le \epsilon \le 0.55$ while 1B is drawn for the ellipticity $0.07 \le \epsilon \le 0.17$.}
\end{figure}

\begin{figure}
\includegraphics[width=8.0cm]{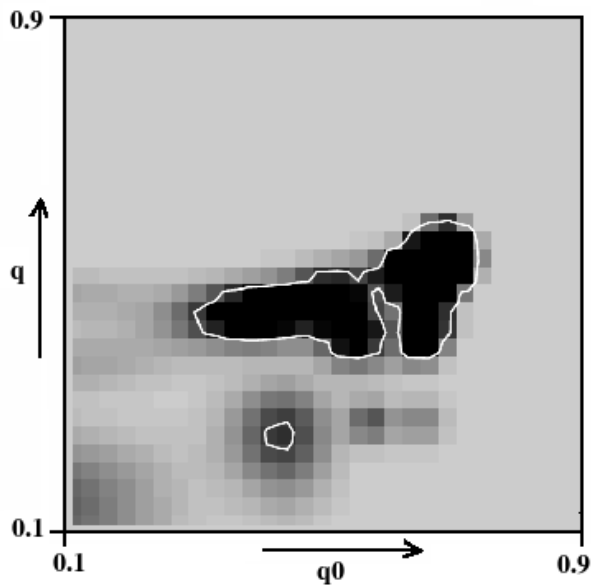}
\caption{\small Plot of marginal posterior density $({\cal{P}})$ as a function of $q_{0}, q_{\infty}(=q)$, summed over various values of $(T_{0}, T_{\infty})$, for NGC 720 using the limits $0.1$ to $0.9$, both for $q_{0}$ and $q_{\infty}$.}
\end{figure}
\begin{figure}
\includegraphics[width=8.0cm]{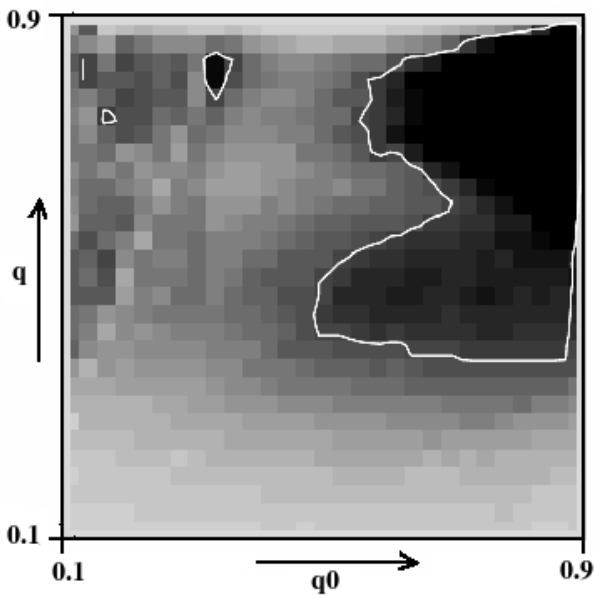}
\caption{\small Same as Fig 2, for NGC 3379.}
\end{figure}
 
Fig. 2 presents the marginal posterior density ${\cal{P}}$ as a function of $(q_{0}, q_{\infty})$, summed over various values of $(T_{0}, T_{\infty})$ for $NGC720$. We use the observational data, as shown in Table 1. We choose the values of both $q_{0}$ and $q_{\infty}$ from the region $0.1 \le (q_{0}, q_{\infty}) \le 0.9$. We use the $M^{2}$ models with $\beta = 1.0$. The probability of the shape is plotted in dark grey shade : darker is the shade, higher is the probability. The white contour encloses the region of $68\%$ highest posterior density (HPD), which may be interpreted as $1 \sigma$ error bar. This figure indicates that higher probability region is confined, approximately between $0.3$ to $0.8$ of $(q_{0}, q_{\infty})$. Hence, it is more appropriate to choose the lower and the upper limits of both $q_{0}$ and $q_{\infty}$ as $0.3$ and $0.8$ for the shape estimates of very flat galaxies. This is discussed further, in sect. 3.1.

For the same choice of the values of $(q_{0}, q_{\infty})$, fig. 3 shows shape ${\cal{P}}$$(q_{0}, q_{\infty})$ of a rounder galaxy $NGC 3379$. We use the observational data $\epsilon_{in} = 0.078, \epsilon_{out} = 0.133$ at $R_{in} = 15^{''}.7$ and $R_{out} = 49^{''}.3$. The effective radius $R_{e}$ of $NGC 3379$ is $37^{''}.5$. We find that the $HPD$ region is confined between$(q_{0}, q_{\infty}) \ge 0.4$ and the highest values of $(q_{0}, q_{\infty})$ allowed in this plot.

\subsection{$\mathbf{NGC 720}$}
The observed data of $NGC720$, is taken from R-band surface photometry of \citet{b13}. The ellipticity $\epsilon$ increases monotonically from $0.315$ at $R_{in} = 8.5$ arcsec to $0.442$ at $R_{out} = 51.8$ arcsec. In this range, the position angle decreases by $3^{o}.5$. We consider the uncertainty in the ellipticity as $0.02$ and in the position angle is $1^{o}.0$, both at $R_{in}$ and at $R_{out}$. These are the typical errors in observations \citep{b8,b14}. The effective radius of the galaxy is $52.0$ arcsec. We use the ensemble of models, as described in sect. 2, with $\beta = 5.0, 2.5, 1.0, 0.5$ and $0.2$. Taking the sum of the marginal posterior density over all possible values of $T_{0}$ and $T_{\infty}$, and taking the unweighted sum over all the models, we obtain shape estimate ${\cal{P}}$ as a function of $(q_{0}, q_{\infty})$.

\begin{figure}
\includegraphics[width=8.0cm]{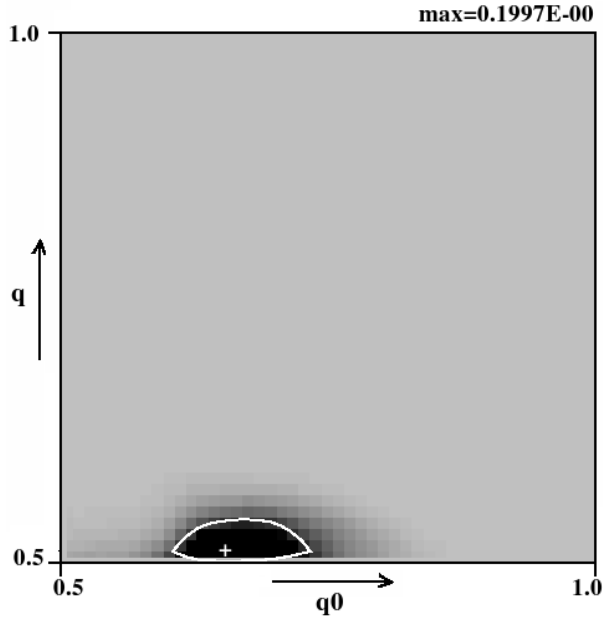}
\caption{\small Plot of unweighted sum of MPD $({\cal{P}})$ as a function of $q_{0}, q_{\infty}(=q)$, for NGC 720 using the limits $0.5$ to $1.0$ both for $q_{0}$ and $q_{\infty}$. The sum is taken over the $M^{2}$ models with $\beta = 5.0, 2.5, 1.0, 0.5$ and $0.2$. Plus marks the location of the maximum probability.}
\end{figure}

\begin{figure}
\includegraphics[width=8.0cm]{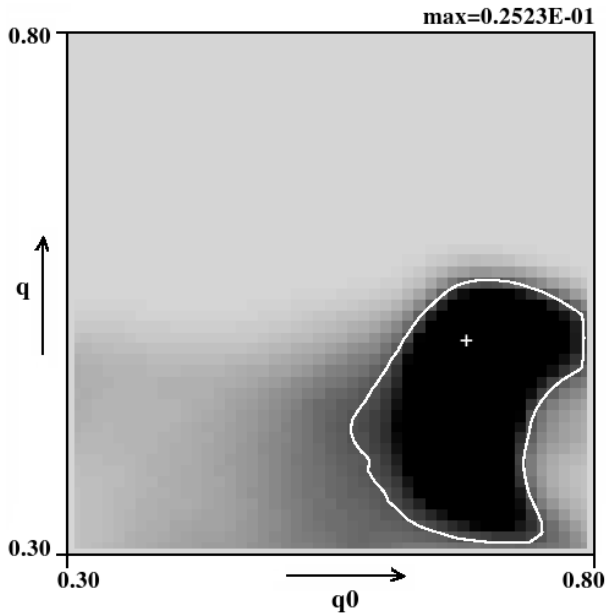}
\caption{\small Same as Fig 4, for NGC 720 using the limits $0.3$ to $0.8$, both for $q_{0}$ and $q_{\infty}$.}
\end{figure}

\begin{figure}
\includegraphics[width=8.0cm]{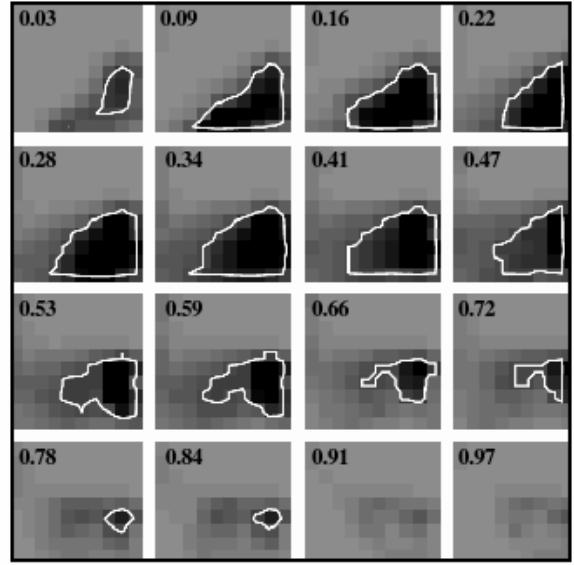}
\caption{\small Three dimensional plot of the unweighted sum of MPD $({\cal{P}})$ as a function of $q_{0}, q_{\infty}, |T_{d}|$, for NGC 720. The sum is taken over the $M^{2}$ models with $\beta = 5.0, 2.5, 1.0, 0.5 $ and $0.2$. Values of $|T_{d}|$ are constant in each section. In each section, $q_{0}$ goes from the left to right hand side from 0.25 to 0.75, and $q_{\infty}$ runs between the same values from the bottom to the top.}
\end{figure}

Fig. 4 presents the shape estimate ${\cal{P}}$ $(q_{0}, q_{\infty})$ of $NGC720$, wherein we have allowed the limits $0.5$ to $1.0$ both for $q_{0}$ and $q_{\infty}$. We find that the $1 \sigma$ region is very narrow which should be the consequence of the choice of the limits of $q_{0}$ and $q_{\infty}$. Examining this limit in fig. 1, we find that this choice falls in the region where high values of the ellipticity will not be reproduced. Therefore, we need to go to smaller values of $(q_{0}, q_{\infty})$ to obtain higher ellipticities, which may be close to observed ellipticities of $NGC720$.

Fig. 5 presents shape ${\cal{P}}$ $(q_{0}, q_{\infty})$ wherein we have allowed the limits $0.3$ and $0.8$ for $q_{0}$ and $q_{\infty}$. $1 \sigma$ region is wider now (but narrow enough to satisfy the requirement of the likelihood dominated shape estimate). Although, it is the plot of $MPD$ $({\cal{P}})$ as a function of shape parameters, which constitute the Bayesian estimate of the shape, some statistical summary of the shape is very convenient for its description. The expectation values $<q_{0}>, <q_{\infty}>$ and location of the peak values $q_{0P}, q_{{\infty}P}$ are such quantities. Table 2 provides such a summary. The expectation values of the flattening at small and at large radii are $<q_{0}> = 0.64$ and $<q_{\infty}> = 0.43$, respectively. 

Both in fig. 4 and 5, we choose the interval between higher and lower limits of $(q_{0}, q_{\infty})$ as $0.5$. This is basically to save the computer time, but maintaining the reliability of the results. Shape calculation requires a very large number of projections, which need to be calculated numerically. We divide the parameter space of $(q_{0}, q_{\infty})$ in $48 \times 48$ square bins of equal size and calculate the likelihood at the centre of each bin. The bin size is small enough, so that the calculated likelihood can be regarded as a continuum function of $(q_{0}, q_{\infty})$, and at the same time, the number of bins is small enough, so that the computer time is not unmanageable.

Fig. 6 shows the 3-dimensional intrinsic shape of $NGC720$ as a function of $q_{0}, q_{\infty}$ and $|T_{d}|$. We cut a total of 16 sections, each perpendicular to $|T_{d}|$ axis, and arrange these sections in a form of a two-dimensional array. The value of $|T_{d}|$ is constant in each section, and is shown in the plot. We find that the $1 \sigma$ region occupies larger area in the sections with smaller values of $|T_{d}|$. Further, in each section of constant $|T_{d}|$, $1 \sigma$ region occupies a small area of $(q_{0}, q_{\infty})$ plane. We find that higher ${\cal{P}}$ is concentrated in sections with $|T_{d}|$ between $0.28$ to $0.47$. The expectation value of $<|T_{d}|> = 0.41$.

\subsection{Intrinsic shapes of $\mathbf{NGC2768}$ and $\mathbf{3605}$}
The observational data used in the models of these galaxies are presented in Table 1. Here also, the data is obtained from R - band surface photometry of \citet{b13}.

\begin{table}
\caption{Observational data of the galaxies.}
\begin{tabular}{@{}ccccccc@{}}
\hline
Galaxy & $R_{e}$ & $R_{in}$ & $R_{out}$ & $\epsilon_{in}$ & $\epsilon_{out}$ & $\Theta_{d}$ \\ \hline
NGC 720  & 52.0 & 8.5 & 51.8 & 0.315 & 0.442 & -3.5 \\
NGC 2768 & 76.5 & 15.8 & 95.4 & 0.364 & 0.569 & -1.4 \\
NGC 3605 & 22.5 & 5.9 & 20.4 & 0.305 & 0.418 & -2.0 \\ \hline \hline
\end{tabular}
\end{table}

Fig. 7 and 8 present the plot of ${\cal{P}}$ of $NGC2768$ and $3605$ as functions of $(q_{0}, q_{\infty})$. The lower and upper limits of $q_{0}$ and $q_{\infty}$ are taken as $0.25$ to $0.75$. The HPD region shows that these galaxies are very flat. The expectation values are $<q_{0}> = 0.63$ and $<q_{\infty}> = 0.32$ for $NGC2768$ and are $<q_{0}> = 0.62$ and $<q_{\infty}> = 0.42$ for $NGC3605$. We find that these galaxies are also intrinsically very flat. Fig. 9 and 10 present the 3-dimensional plot of $\cal{P}$ of  $NGC2768$ and $3605$ as a function of $q_{0}$, $q_{\infty}$ and $|T_{d}|$.

Statistical summary of the intrinsic shapes of all the three flat galaxies $NGC 720, 3605$ and $2768$ is presented in Table 3. Here, the values are taken from the 3-dimensional shape estimates. The expected and the peak values of $q_{0}$ and $q_{\infty}$ as obtained from 2-dimensional estimates ${\cal{P}}$ $(q_{0}, q_{\infty})$ are reported in Table 2. These values are quite close but not exactly the same as those reported in Table 3. The differences may be attributed to "resolution" : in 2-dimensional shape estimates, we have divided the parameters space of $(q_{0}, q_{\infty})$ in $48 \times 48$ divisions, whereas in 3-dimensional shape estimate, space $(q_{0}, q_{\infty})$ is divided in $10 \times 10$ divisions for each $|T_{d}|$.

\begin{figure}
\includegraphics[width=8.0cm]{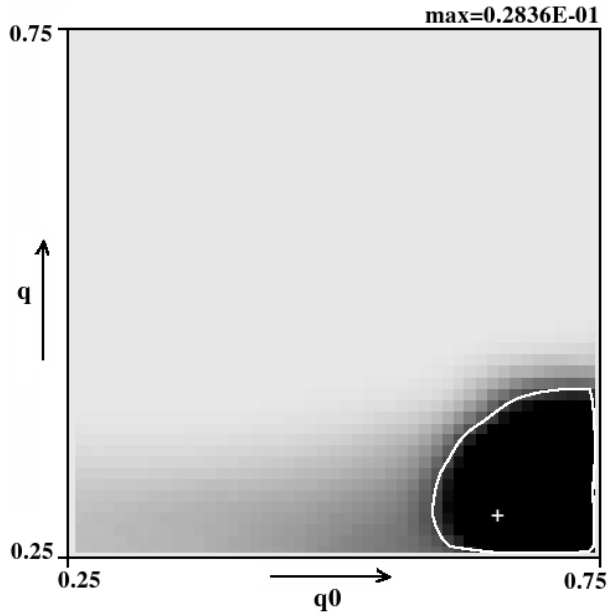}
\caption{\small Same as Fig 4, for NGC 2768 using the limits $0.25$ to $0.75$, both for $q_{0}$ and $q_{\infty}$.}
\end{figure}

\begin{figure}
\includegraphics[width=8.0cm]{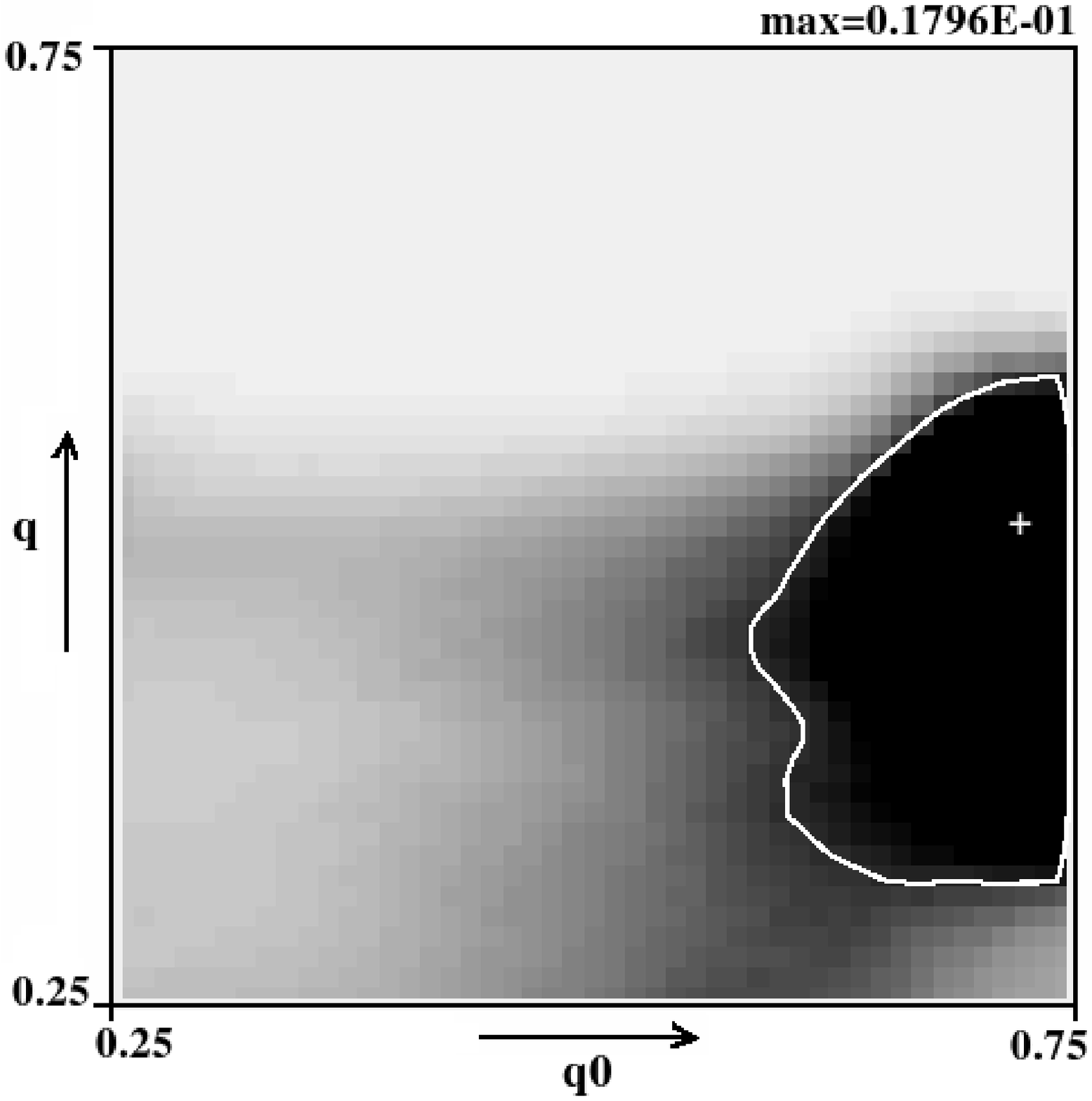}
\caption{\small Same as Fig 4, for $NGC 3605$ using the limits $0.25$ to $0.75$, both for $q_{0}$ and $q_{\infty}$.}
\end{figure}

\begin{figure}
\includegraphics[width=8.0cm]{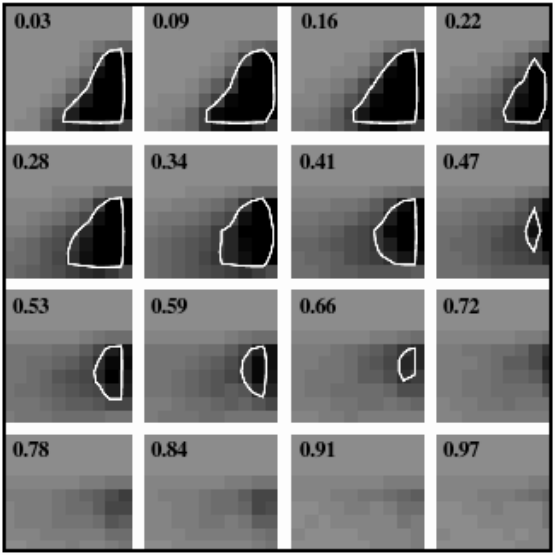}
\caption{\small Same as Fig 6, for $NGC 2768$.}
\end{figure}
\begin{figure}
\includegraphics[width=8.0cm]{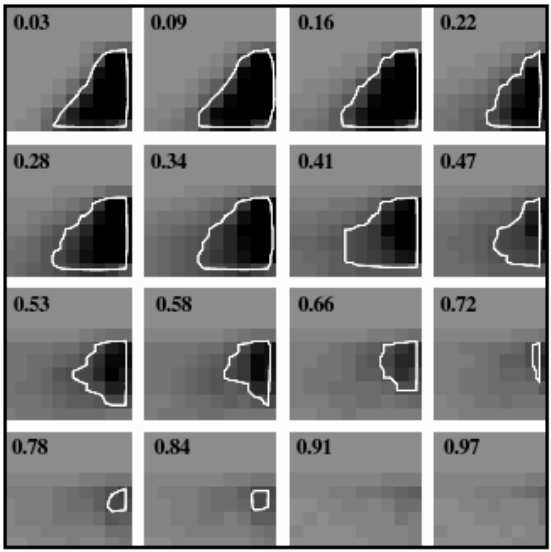}
\caption{\small Same as Fig 6, for $NGC 3605$.}
\end{figure}

\begin{table}
\caption{Statistical summary of the 2-dimensional shape estimates ${\cal{P}}(q_{0}, q_{\infty})$ of the galaxies.}
\begin{tabular}{@{}lcccc@{}}
\hline
Galaxy & $q_{0p}$ & $q_{\infty p}$ & $<q_{0}>$ & $<q_{\infty}>$ \\ \hline
$NGC 720$  &  0.68 &  0.48 &  0.64 &  0.43 \\
$NGC 2768$ &  0.65 &  0.29 &  0.63 &  0.32 \\
$NGC 3605$ &  0.72 &  0.49 &  0.62 &  0.42 \\ \hline \hline
\end{tabular}
\end{table}
\begin{table}
\caption{Statistical summary of the 3-dimensional shape estimates ${\cal{P}}(q_{0}, q_{\infty}, |T_{d}|)$ of the galaxies.}
\begin{tabular}{@{}lcccccc@{}}
\hline
Galaxy & $q_{0p}$ & $q_{\infty p}$ & $|T_{dp}|$ & $<q_{0}>$ & $<q_{\infty}>$ & $<|T_{d}|>$ \\ \hline
$NGC 720$  &  0.68 &  0.38 &  0.41 &  0.56 &  0.40 &  0.41  \\
$NGC 2768$ &  0.68 &  0.28 &  0.22 &  0.62 &  0.33 &  0.39  \\ 
$NGC 3605$ &  0.68 &  0.43 &  0.16 &  0.60 &  0.41 &  0.37  \\ \hline \hline
\end{tabular}
\end{table}

\section{Results and discussion}

We have presented the intrinsic shapes of 3 very flat galaxies. A specific feature of these estimates is the choice of the lower limit of $q_{0}$ and $q_{\infty}$. The lower limit was chosen as $0.5$ by $C08$ and as $0.4$ by \citet{b17,b18} in their investigation of galaxies which are comparatively rounder, without mentioning any specific reason for this choice. Through the plots of fig. 1, and the detail discussion of the plots of ${\cal{P}}$ for $NGC 720$, we justified the choice of very small values of $(q_{0}, q_{\infty})$ as their lower limits for very flat elliptical galaxies. We took the lower limit as either $0.25$ or $0.3$.

A summary of the intrinsic shapes of these very flat galaxies is presented in Table 2 and Table 3. We find that these galaxies are little rounder inside (average value of $<q_{0}> \sim 0.6$), but very flat outside (average value of $<q_{\infty}> \sim 0.4$). Following the nomenclature introduced in $C08$, these galaxies may be termed as RF type. 

Intrinsic shapes of elliptical galaxies have implications for their formation and evolution. As the galaxies studied here are very flat, we have given emphasis on the shape ${\cal{P}}(q_{0}, q_{\infty})$, and on informations about the flattening at inner and at outer radii. 

\section*{Acknowledgements}
DKC would like to thank the coordinator, IUCAA reference centre, Pt. Ravishankar Shukla University, for the technical support. AKD gratefully acknowledges the award of Rajiv Gandhi National Fellowship grant No. F-16-71/2006 (SA-II), from UGC, New Delhi, India. We thank the reviewer for his comments which helped us to improve the paper in its present form.

\appendix
\section{Very flat de Zeeuw - Carollo models}
A simple family of triaxial models, with ellipticity variation and position angle twist was presented by \citet{b10} with density distribution
\begin{eqnarray}
\rho (r, \theta, \phi) = f(r)-g(r) Y^{0}_{2}(\theta) + h(r) Y^{2}_{0}(\theta, \phi)\ , 
\end{eqnarray}
where $f(r)$ is same as $(1)$, $g(r)$ and $h(r)$ are two radial functions, and $Y^{0}_{2}$ and $Y^{2}_{2}$ are the usual spherical harmonics. Here, $(r, \theta, \phi)$ are the standard polar co-ordinates. The projected surface density of $(A1)$ can be calculated easily, and often analytically. $g(r)$ and $h(r)$ are fixed by assiging axial ratios $(p_{0}, q_{0})$ and $(p_{\infty}, q_{\infty})$ respectively, at small large radii, where constant - $\rho$ surfaces are approximately ellipsoidal. Numerical distribution function was shown to exist for prolate triaxials : $ (p,q) = (0.65, 0.60)$ and for oblate triaxials : $ (p,q) = (0.95, 0.65)$, wherein it is assumed that $q_{0} = q_{\infty} =q, p_{0} = p_{\infty} =p$ and $\gamma = 1.0$ or $1.5$ \citep{b24}. The rounder versions of these models are employed successfully in many investigations, including the shape estimates (\citealt{b23}; $C08$).

However, very flat versions of de Zeeuw - Carollo models have several undesirable features. It was realized by \citet{b10} that the constant $\rho$ surfaces become peanut shaped or dimpled for large values of flattening. Such models can not be a "true" representation of the shape of an elliptical galaxy.

In addition to above, we now find the appearance of narrow regions, where $\rho$ is negative. Clearly, it is unphysical to call such negative $\rho$ as mass density. Such negative $\rho$ appears in polar regions $(\theta \sim 0^{o}.0)$, at an intermediate $r$ extending from some $r_{low}$ to $r_{high}$. Tables $(A1)$ and $(A2)$ show some of the regions of negative $\rho$ on $\phi = \frac{\pi}{2}$ plane.

Triaxials models of similar form as $(A1)$ was proposed by \citet{b15} as a numerical model, where $f(r)$ is taken as the modified Hubble density distribution. Later, it was put into an analytical form by \citet{b11}. Projected properties of such triaxial modified Hubble model was studied by \citet{b6}. We now find that sufficiently flat versions of triaxial modified Hubble model also exhibit regions of negative $\rho$.

We find that the appearance of negative $\rho$ regions is correlated with the dimpleness of constant $\rho$ surfaces. We examine an extended version of models $(A1)$, which includes terms with high order spherical harmonics $Y^{0}_{4}, Y^{2}_{4}$ and $Y^{4}_{4}$. Such models were studied by \citet{b4}. It was found that the dimpleness reduces i.e., the models become more ellipsoidal - like. We now find that for the same choice of the parameters as in Table $(A1)$, the interval $(r_{high} - r_{low})$ of negative $\rho$ decreases.

It will be interesting to extend the studies of particle orbit and the numerical distribution function of \citet{b24}, to the models which exhibit regions of negative $\rho$.
\begin{table}
\caption{Regions of negative $\rho$.}
\begin{tabular}{@{}ccc@{}}
\hline
\multicolumn{3}{c}{$p = p_{0} = p_{\infty} = 0.9, \gamma = 1.5, \theta = 0^{o}.0, \phi = \frac{\pi}{2}$.} \\ \hline
$q = q_{0} = q_{\infty}$ & $\frac{r_{low}}{b}$ & $\frac{r_{high}}{b}$ \\ \hline
0.55 & 1.93 & $>$ 6.40 \\
0.56 & 2.31 & 6.27 \\
0.57 & 3.08 & 4.48 \\ \hline
$\ge$ 0.58 & \multicolumn{2}{l}{ $\rho$ is positive at all $r$} \\\hline \hline
\end{tabular}
\end{table}
\begin{table}
\caption{Regions of negative $\rho$.}
\begin{tabular}{@{}lccccccc@{}}
\hline
\multicolumn{3}{c}{$p = p_{0} = p_{\infty} = 0.9, \gamma = 1.5, \phi = \frac{\pi}{2}, q = 0.57 $} \\ \hline
$\theta$ & $\frac{r_{low}}{b}$ & $\frac{r_{high}}{b}$ \\ \hline
$0^{o}.0$ & 3.08 & 4.48 \\
$1^{o}.0$ & 3.08 & 4.48 \\
$2^{o}.0$ & 3.21 & 4.36 \\
$3^{o}.0$ & 3.46 & 3.85 \\ \hline
$\ge$ $4^{o}.0$ & \multicolumn{2}{l}{ $\rho$ is positive at all $r$} \\ \hline \hline
\end{tabular}
\end{table}

\bsp

\label{lastpage}

\end{document}